\documentclass[fleqn,usenatbib]{mnras}

\usepackage{newtxtext,newtxmath}

\usepackage[T1]{fontenc}

\DeclareRobustCommand{\VAN}[3]{#2}
\let\VANthebibliography\thebibliography
\def\thebibliography{\DeclareRobustCommand{\VAN}[3]{##3}\VANthebibliography}

\usepackage{graphicx}	
\usepackage{amsmath}	
\usepackage{orcidlink}

\title[Delivery of COMs to the system of Jupiter]{Delivery of complex organic molecules to the system of Jupiter}

\author[T. Benest Couzinou et al.]{
Tom Benest Couzinou$^{\orcidlink{0000-0002-8719-7867}}$$^{1}$\thanks{E-mail: tommy.benest@gmail.com},
Alizée Amsler Moulanier$^{\orcidlink{0009-0000-7492-1476}}$$^{1}$,
and Olivier Mousis$^{\orcidlink{0000-0001-5323-6453}}$$^{2}$
\\
$^{1}$Aix-Marseille Universit\'e, CNRS, CNES, Institut Origines, LAM, Marseille, France\\
$^{2}$Solar System Science and Exploration Division, Southwest Research Institute, 1301 Walnut St, Ste 400, Boulder, CO, USA
}

\date{Accepted XXX. Received YYY; in original form ZZZ}

\pubyear{2025}

\begin{document}
\label{firstpage}
\pagerange{\pageref{firstpage}--\pageref{lastpage}}
\maketitle

\begin{abstract}
Complex organic molecules are key markers of molecular diversity, and their formation conditions in protoplanetary disks remain an active area of research. These molecules have been detected on a variety of celestial bodies, including icy moons, and may play a crucial role in shaping the current composition of the Galilean moons. Experimental studies suggest that their formation could result from UV irradiation or thermal processing of NH$_3$:CO$_2$ ices. In this context, we investigate the formation of complex organic molecules in the protosolar nebula and their subsequent transport to the Jupiter system region. Lagrangian transport and irradiation simulations of 500 individual particles are performed using a two-dimensional disk evolution model. Based on experiments with UV irradiation and thermal processing of CO$_2$:NH$_3$ ice, this model allows us to estimate the estimate the potential for the formation of complex organic molecules through these processes. Almost none of the particles released at a local temperature of 20 K (corresponding to $\sim$12 AU from the Sun) reach the location of the system of Jupiter. However, when released at a local temperature of 80 K ($\sim$7 AU), approximately 45\% of the centimetric particles and 30\% of the micrometric particles can form complex organic molecules via thermal processing, subsequently reaching the location of the system of Jupiter within 300 kyr. Assuming that the Galilean moons formed in a cold circumplanetary disk around Jupiter, the nitrogen--bearing species potentially present in their interiors could have originated from the formation of complex organic molecules in the protosolar nebula.
\end{abstract}




\section{Introduction} \label{sec:intro}

Europa, Ganymede, and Callisto are believed to harbor subsurface oceans beneath their icy crusts, while Io is thought to possess a liquid magma ocean beneath its surface \citep{Khurana1998, Pappalardo1999, Kivelson2000, Kivelson2002, Khurana2011}. These subsurface environments raise critical questions about the potential habitability of the moons and the need for a deeper understanding of their internal composition. 

A key uncertainty is the role of complex organic molecules (COMs) in shaping the moons' composition over time. While multiple studies have investigated their presence in the Galilean moons \citep{Delitsky_Lane_1998,McCord_1998}, direct evidence remains elusive, with detections so far limited to the plumes of Saturn's icy moon Enceladus \citep{Postberg2008,Postberg2018,WaiteJr2009}.
In addition, elements such as nitrogen, oxygen, carbon, and sulfur, potentially introduced by COMs during moon formation, add complexity to their chemical evolution.
COMs, comprising multiple carbon, hydrogen, oxygen, and possibly nitrogen atoms \citep{belloche_increased_2009, tenelanda-osorio_effect_2022}, have been detected in comets and star-forming regions \citep{briggs_comet_1992, bisschop_testing_2007, bergantini_formation_2017, martin-domenech_uv_2018}. 
Laboratory experiments demonstrate that ices such as CO, CO$_2$, NH$_3$, H$_2$O, and CH$_3$OH can produce diverse molecules through UV irradiation \citep{bossa_carbamic_2008,tenelanda-osorio_effect_2022}, ion bombardment \citep{pilling_radiolysis_2010,kanuchova_synthesis_2016}, electron bombardment \citep{lafosse_reactivity_2006,esmaili_glycine_2018}, and thermal processing \citep{bossa_carbamic_2008,theule_thermal_2013}. These findings offer crucial insights into the mechanisms driving molecular complexity in the dynamic environments of protoplanetary disks (PPDs).

Understanding molecular complexity in PPDs is crucial for exploring the chemical environments of circumplanetary disks (CPDs), where moons around giant planets are thought to form \citep{Canup_Ward_2002, Canup_Ward_2006, Sasaki_Stewart_Ida_2010}. CPDs share structural and chemical similarities with PPDs, influencing the molecular composition of the material they accrete.  The nature of CPDs, whether massive \citep{Lunine_Stevenson_1982, keith_accretion_2014, Szulágyi_Masset_Lega_Crida_Morbidelli_Guillot_2016, Schneeberger2025} or low-mass \citep{Canup_Ward_2002, Canup_Ward_2006, Coradini_Magni_Turrini_2010, Ronnet_Johansen_2020, Anderson_Mousis_Ronnet_2021, Mousis_Schneeberger_Lunine_Glein_Bouquet_Vance_2023}, significantly affects their properties. In low-mass CPDs, colder environments lack sufficient thermal energy to vaporize incoming material, allowing the molecular composition of particles from the PPD to remain largely unchanged \citep{Canup_Ward_2002, Canup_Ward_2006, Mo04, Mo09}.

The cold formation scenario for the Galilean moons suggests that their bulk compositions primarily reflect materials inherited from the protosolar nebula (PSN) \citep{Lubow1999,Canup_Ward_2002,Tanigawa2012,Christiaens2019,Szulágyi2022}. In this framework, COMs formed within the PSN were delivered to the CPD, where they likely served, along with other molecular species, as fundamental building blocks for the moons.
This study investigates the delivery mechanisms of COMs to the system of Jupiter -- and by extension, to its Galilean moons -- assuming that these molecules serve as the primary source of nitrogen. As the baseline of this study, we use the laboratory experiments simulating COM formation in NH$_3$:CO$_2$ ices subjected to irradiation and thermal processing \citep{bossa_carbamic_2008}. These experiments determine the irradiation doses necessary for COM formation and the thermal energy thresholds required for their synthesis during heating. By comparing these experimental results with conditions in the PSN, the study examines how COMs form, persist, and are transported within the disk before being delivered to the Jupiter system region. The ultimate goal is to understand how COMs could act as precursors to moons formation in a cold CPD environment, and contribute to the volatile inventory and molecular diversity observed on the Galilean moons. To achieve this, we use a time-dependent PPD model, as described in \cite{Aguichine_Mousis_Devouard_Ronnet_2020} and \cite{Schneeberger_Mousis_Aguichine_Lunine_2023}, which employs a diffusion equation in the Eulerian framework to simulate the evolution of the PSN. Additionally, the radial and vertical motion of particles within the disk is modeled using a Lagrangian scheme \citep{Ciesla_2010, Ciesla_2011, benest2024}. By evaluating the temperatures and irradiation doses experienced by the particles and comparing them with experimental data, we assess the formation and transport of COMs within the PSN and their eventual delivery to the Jupiter system region and thus to Jupiter's CPD.

The paper is organized as follows. In Section \ref{sec:method}, we detail the parameters and assumptions of the PSN model and outline the experimental framework established by \cite{bossa_carbamic_2008}, which serves as the basis for our model. Section \ref{sec:results} compares the results of our numerical simulations with the experimental data. Finally, Section \ref{sec:discussion} examines the broader implications of our findings for the transport and survival of complex organic molecules in the protosolar nebula and concludes the study.

\section{Methodology} \label{sec:method}

This section describes the modules used to compute the evolution of the disk, the transport and irradiation of particles, as well as the experimental conditions upon which our results are based.

\subsection{Evolution of the disk}

We utilize a one-dimensional, time-dependent accretion disk model, employing the formalism of $\alpha$-turbulent viscosity, expressed as $\nu = \alpha c_{\mathrm{s}}^2 / \Omega_{\mathrm{K}}$ \citep{Shakura_Sunyaev_1973}. Here, $\alpha$ is the viscosity parameter set to $10^{-3}$ \citep{Hartmann_Calvet_Gullbring_D’Alessio_1998,Hueso_Guillot_2005,Desch_Estrada_Kalyaan_Cuzzi_2017}, and
$\Omega_{\mathrm{K}} = \sqrt[]{GM_{\mathrm{\odot}} / r^3}$ is the Keplerian frequency, where $M_{\mathrm{\odot}}$ is the mass of the star, $r$ is the distance from the star, and $G$ stands for the gravitational constant. The isothermal sound speed, $c_{\mathrm{s}} = \sqrt[]{R_{\mathrm{g}} T_{\mathrm{m}} / \mu}$, is a function of the ideal gas constant $R_{\mathrm{g}}$, the midplane temperature of the disk $T_{\mathrm{m}}$, and the mean molar mass of the gas $\mu = 2.31$ g.mol$^{-1}$.

The evolution of the disk is governed by variations in its gas surface density, $\Sigma_{\mathrm{g}}$, under the assumption of vertical hydrostatic equilibrium, and its midplane temperature, $T_\mathrm{m}$. We compute $\Sigma_{\mathrm{g}}$ using the following differential equation \citep{Lynden-Bell_Pringle_1974}:

\begin{equation} \label{eq:dSigmaG/dt}
\frac{\partial \Sigma_{\mathrm{g}}}{\partial t} = \frac{3}{r} \frac{\partial }{\partial r} \left [r^{ \frac{1}{2} } \frac{\partial }{\partial r} \left ( r^{ \frac{1}{2} } \Sigma_{\mathrm{g}} \nu \right ) \right ].
\end{equation}

\noindent $T_\mathrm{m}$ is calculated by accounting for all heating sources \citep{Hueso_Guillot_2005}:

\begin{equation} 
\label{eq:temperature}
\begin{split}
T_\mathrm{m}^4 & = \frac{1}{2 \sigma_\mathrm{sb}} \left ( \frac{3}{8} \tau_\textup{R} + \frac{1}{2 \tau_\textup{P}} \right ) \Sigma_\mathrm{g} \nu \Omega_\mathrm{K}^2  + T_\mathrm{amb}^4,
\end{split}
\end{equation}

\noindent where $\sigma_\mathrm{sb}$ is the Boltzmann constant. $\tau _\textup{R} = \frac{\kappa_\textup{R} \Sigma_\mathrm{g}}{2}$ is the Rosseland optical depth and $\tau _\textup{P} = 2.4 \tau _\textup{R}$ is the Planck optical depth, where $\kappa_\textup{R}$ represents the mean Rosseland opacity \citep{Bell_Lin_1994}. Finally $T_\mathrm{amb} = 10~K$ is the interstellar medium temperature in dark molecular clouds. Assuming the disk is vertically isothermal \citep{Ciesla_2010,Ronnet_Mousis_Vernazza_2017,benest2024}, we can derive its vertical density $\rho_{\mathrm{g}}$ as follows:

\begin{equation} \label{eq:vertical_structure}
\rho_{\mathrm{g}}(r,z) = \rho_{\mathrm{m}}(r) e^{-\frac{z^2}{2H^2}}
\end{equation}

\noindent with $\rho_{\mathrm{m}}$ corresponding to the midplane density,  $H = c_{\mathrm{s}} / \Omega_{\mathrm{K}}$ the disk scale height and $z$ to the vertical position. For a detailed description of the accretion disk model, we refer the reader to \cite{Aguichine_Mousis_Devouard_Ronnet_2020}, \cite{Schneeberger_Mousis_Aguichine_Lunine_2023} and \cite{benest2024}.

\subsection{Transport and irradiation of particles}

The vertical trajectories of particles are determined by solving the following differential equation \citep{Dubrulle_Morfill_Sterzik_1995, Gail_2001, Ciesla_2010}:

\begin{equation} \label{eq:euler,z}
z_i = z_{i-1} + v_{\mathrm{eff},z} \delta t + R_1 \left [ \frac{2}{\sigma^2}D(z)\delta t \right ]^{\frac{1}{2}},
\end{equation}

\noindent where z$_{i-1}$ and $z_{i}$ are the initial vertical position of the particle, and the position after a given time step $\delta t = \frac{1}{\Omega_K}$, respectively. $D = \frac{\nu}{1+\mathit{St}^2}$ is the diffusivity, $St$ is the Stokes number, and $R_1 \in [-1;1] $ is a random number, with $\sigma^2$ its distribution variance (= 1/3 for uniform distribution), taking into account the random movement due to the viscous effects from eddies in the turbulent fluid \citep{Visser_1997}, using a Monte-Carlo approach \citep{Ciesla_2010}.

The advection velocity term $v_{\mathrm{eff},z}$ consists of three components \citep{Ciesla_2010,Ronnet_Mousis_Vernazza_2017,Mousis_Ronnet_Lunine_Maggiolo_Wurz_Danger_Bouquet_2018,benest2024}:
\begin{equation} \label{eq:v_eff,z}
v_{\mathrm{eff},z}= -t_s \Omega_K^2 z + \frac{D}{\rho_\mathrm{g}} \frac{\partial \rho_\mathrm{g}}{\partial z} + \frac{\partial D}{\partial z}.
\end{equation}

\noindent The first term is given by the terminal velocities of the particles, accounting for drag forces \citep{Cuzzi_Weidenschilling_2006,Ciesla_2010}. The second component arises from the vertical gradient of the nebular gas density through which the species diffuses \citep{Ciesla_2010,Ronnet_Mousis_Vernazza_2017,Mousis_Ronnet_Lunine_Maggiolo_Wurz_Danger_Bouquet_2018}, and the last component considers the vertical gradient of the diffusion coefficient, which is assumed to be zero in a vertically isothermal disk. $t_\mathrm{s}$ is the stopping time, and represents the timescale required for the grains, assumed spherical, to transfer all their angular momentum to the gas. In the Epstein regime, $t_\mathrm{s}$ is described by \citep{Perets_Murray-Clay_2011,Mousis_Ronnet_Lunine_Maggiolo_Wurz_Danger_Bouquet_2018,benest2024}:

\begin{equation} \label{eq:t_s}
t_\mathrm{s} = \frac{ \rho_\mathrm{s} }{\rho_\mathrm{g} } \frac{R_\mathrm{s}}{v_\mathrm{th}},
\end{equation}

\noindent where $v_\mathrm{th} = \sqrt[]{8/\pi} c_\mathrm{s} $ is the thermal sound velocity, $R_s$ is the solid dust grain radius and $\rho_\mathrm{s}$ its density (here 1 g.cm$^{-3}$).

The radial transport equation is split into two analogous parts, considering the Cartesian $x$ and $y$ axis. The $x$ axis equations are briefly described below, and since axial symmetry is assumed, the equations along the $y$ axis are equivalent. The assumption of axial symmetry also justifies ignoring orbital motion \citep{Ciesla_2011}. Following the method applied to the vertical equations along the $z$ axis, the evolution of particles along the $x$ axis is described analogously and can be expressed as \citep{Ciesla_2011}:

\begin{equation} \label{eq:euler,x}
 x_i = x_{i-1} + v_{\mathrm{eff},x} \delta t + R_1 \left [ \frac{2}{\sigma^2}D(x)\delta t \right ]^{\frac{1}{2}}.
\end{equation}

\noindent Here, $R_1$ and $\sigma$ retain their definitions as given in Eq. \ref{eq:euler,z}. The advection velocity term, $v_{\mathrm{eff},x}$, along the $x$ axis is composed of three components \citep{Ciesla_2011}:

\begin{equation} \label{eq:v_eff,x}
v_{\mathrm{eff},x} = v_r \frac{x_{i-1}}{r_{i-1}} + \frac{D}{\rho_\mathrm{g}} \frac{\partial \rho_\mathrm{g}}{\partial r} \frac{x_{i-1}}{r_{i-1}} + \frac{\partial D}{\partial r}\frac{x_{i-1}}{r_{i-1}}
.\end{equation}

\noindent The middle term of Eq. \ref{eq:v_eff,x}, represents the density gradient in the midplane. The right term reflects the gradient of diffusivity along the $x$ axis. Finally, the left term corresponds to the radial velocity of the particles, accounting for radial drift and gas coupling \citep{Lynden-Bell_Pringle_1974,Weidenschilling_1977,Nakagawa_Sekiya_Hayashi_1986,Aguichine_Mousis_Devouard_Ronnet_2020}:

\begin{equation} \label{eq:v_r}
v_r = -\frac{ 1}{1 + \mathrm{St}^2}\frac{3}{2} \frac{\nu}{r} (1+2Q) + \frac{2\mathrm{St} }{1+\mathrm{St}^2}\frac{c_s^2}{r\Omega_K} \frac{\mathrm{dln}P}{\mathrm{dln}r},
\end{equation}

\noindent where $Q = \frac{\mathrm{d}\mathrm{ln}(\Sigma_g \nu)}{\mathrm{d}\mathrm{ln}(r)}$ accounts for the direction of the gas velocity, driven by the centrifugal radius position. The first term of Eq. \ref{eq:v_r} represents the contribution of the gas velocity on coupled particles, while the second term corresponds to the contribution of the gas drag the decoupled particles experience. The relative contributions are determined by the Stokes number of the particles, which depends on their size. For instance, radial drift generally dominates over gas coupling for larger particles.

\begin{figure*}
\centering
\includegraphics[width=0.9\textwidth]{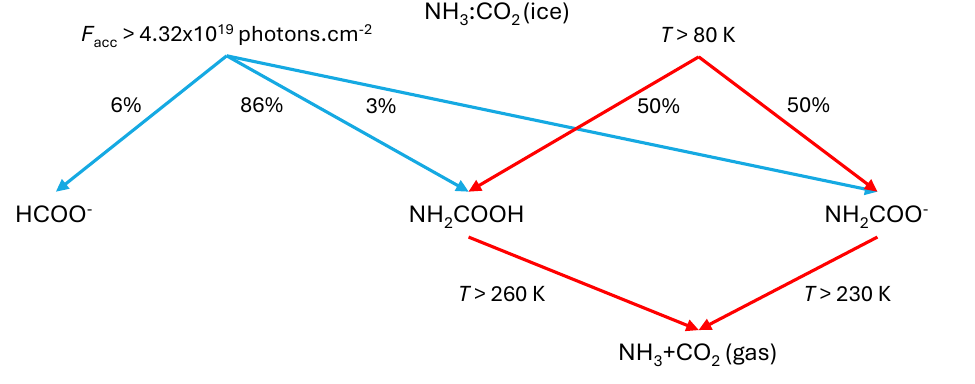}
\caption{Experimental conditions and results adapted from \citet{bossa_carbamic_2008}, which serve as the foundation for our study. The figure depicts the products formed during the processing of NH$_3$:CO$_2$ ices, highlighting reactions driven by thermal processing (in red) and UV irradiation (in blue). Chemical pathways are influenced by the temperature condition $T > 80$ K and an accumulated irradiation dose $F_\mathrm{acc} > 4.32 \times 10^{19}$ photons.cm$^{-2}$. The percentages represent the relative abundances of the newly formed molecules.}
\label{fig:reactions}
\end{figure*}

Along its trajectory, a particle experiences varying levels of irradiation, resulting in an accumulated dose, $F_{\text{acc}}$, calculated as the time-integrated irradiation flux over each timestep. The irradiation flux $F(r,z)$ of UV photons received by a particle in the disk is computed via the following formula \citep{Ciesla_2010,Ciesla_Sandford_2012} :
    
\begin{equation}
F(r,z) = F_0 e^{-\tau (r,z)},
\end{equation}
    
\noindent with
    
\begin{equation} 
\tau (r,z) = \int_{|z|}^{\infty} \rho_{\mathrm{g}} (r,z) \kappa \mathrm{d}z.
\end{equation}
    
\noindent Here $\kappa$ is the mean Rosseland opacity and $\tau (r,z)$ is the optical depth of suspended matter above the particle. Although the opacity adopted here is frequency-averaged, it more accurately captures the thermal structure of the disk across different temperature regimes than a frequency-dependent opacity (see discussion in \citealt{benest2024}). $F_0$ is the incident UV interstellar flux equal to $10^8$ photons cm$^{-2}$ s$^{-1}$ (= 1 G$_0$ = $10^{12}$ photons m$^{-2}$ s$^{-1}$) \citep{Habing_1968,Parravano_Hollenbach_McKee_2000,Yeghikyan_2009,Ciesla_Sandford_2012}.

\subsection{Experimental conditions}

Our modeling is based on the experiment described in \cite{bossa_carbamic_2008}, in which pure CO$_2$ and NH$_3$ were placed in a vacuum chamber at $\sim$ 10$^{-7}$ mbar and 10 K. These ices were heated (at a rate of 4 K$\cdot$min$^{-1}$) or irradiated with UV photons (with wavelengths higher than 120 nm), either individually or mixed at a 1:1 ratio.

During thermal processing, CO$_2$ and NH$_3$ ices sublimate individually at 90--105 K and 100--120 K, respectively, without forming COMs. When mixed in a 1:1 ratio and heated, the reactants begin to react above 80 K and are completely consumed by 130 K. Two products are formed: ammonium carbamate, [NH$_2$COO$^-$][NH$_4^+$] (C), and carbamic acid, NH$_2$COOH (D), with a 1:1 ratio observed at 140 K. Both species sublimate between 230 and 260 K: C decomposes into CO$_2$ and NH$_3$ vapors at 230 K, while D partially decomposes during desorption at 260 K. Only 51\% of the deposited CO$_2$ reacts with NH$_3$, while the remaining fraction sublimates.

Under VUV irradiation of NH$_3$:CO$_2$=1:1 ice at 10 K, an irradiation dose of $4.32\times10^{19}$ photons cm$^{-2}$ consumed 50\% of the reactants, forming carbamic acid NH$_2$COOH, ammonium carbamate [NH$_2$COO$^-$][NH$_4^+$], and ammonium formate [HCOO$^-$][NH$_4^+$]. Intermediate species like HCOOH, HNCO, NH$_4^+$HCOO$^-$, NH$_4^+$OCN$^-$, and COOH formed and undergone reactions. 86$\%$ of the newly formed molecules are carbamic acid, 6\% are HCOO$^-$, and 3\% are ammonium carbamate, resulting in a carbamate:carbamic acid ratio of 1:28. Since \citet{bossa_carbamic_2008} did not report any limitations regarding the effect of temperature on the irradiation-induced chemistry of these ices, we assume in the following that COMs can form if the material experiences this irradiation dose, irrespective of temperature. In reality, however, temperature is likely to influence the efficiency and pathways of the irradiation process. Overall, our results indicate that ice irradiation is more effective in the colder regions of the protoplanetary disk located beyond the formation region of the giant planets. The experimental conditions are summarized in Fig. \ref{fig:reactions}, with additional details provided from \cite{bossa_carbamic_2008}.

The following results consider two potential pathways for the formation of COMs in the PPD. A first aspect explored is whether disk particles could attain temperatures between 80 and 230 K, a range enabling COM formation. A second focus is on whether the accumulated irradiation on these particles might reach or surpass the threshold dose of $4.32 \times 10^{19}$ photons cm$^{-2}$, which is also required for COM synthesis. Although we have seen that not all reactants react and form COM immediately after irradiation or heat treatment, our simulation considers particles that meet the conditions to always undergo COM formation.

\section{Results} \label{sec:results}

This section presents calculations on the migration and irradiation of grains under the specified experimental conditions. To investigate the influence of particle size, the analysis examines the migration of grains of two different sizes: 1 cm and 1 $\mu$m. Simulations are performed for two initial locations in the disk, 7 AU and 12 AU corresponding to a local temperature of 80 and 20 K, respectively. Each scenario includes 500 independent simulations of NH$_3$:CO$_2$ icy particles.

\begin{figure*}
\centering
\includegraphics[width=\textwidth]{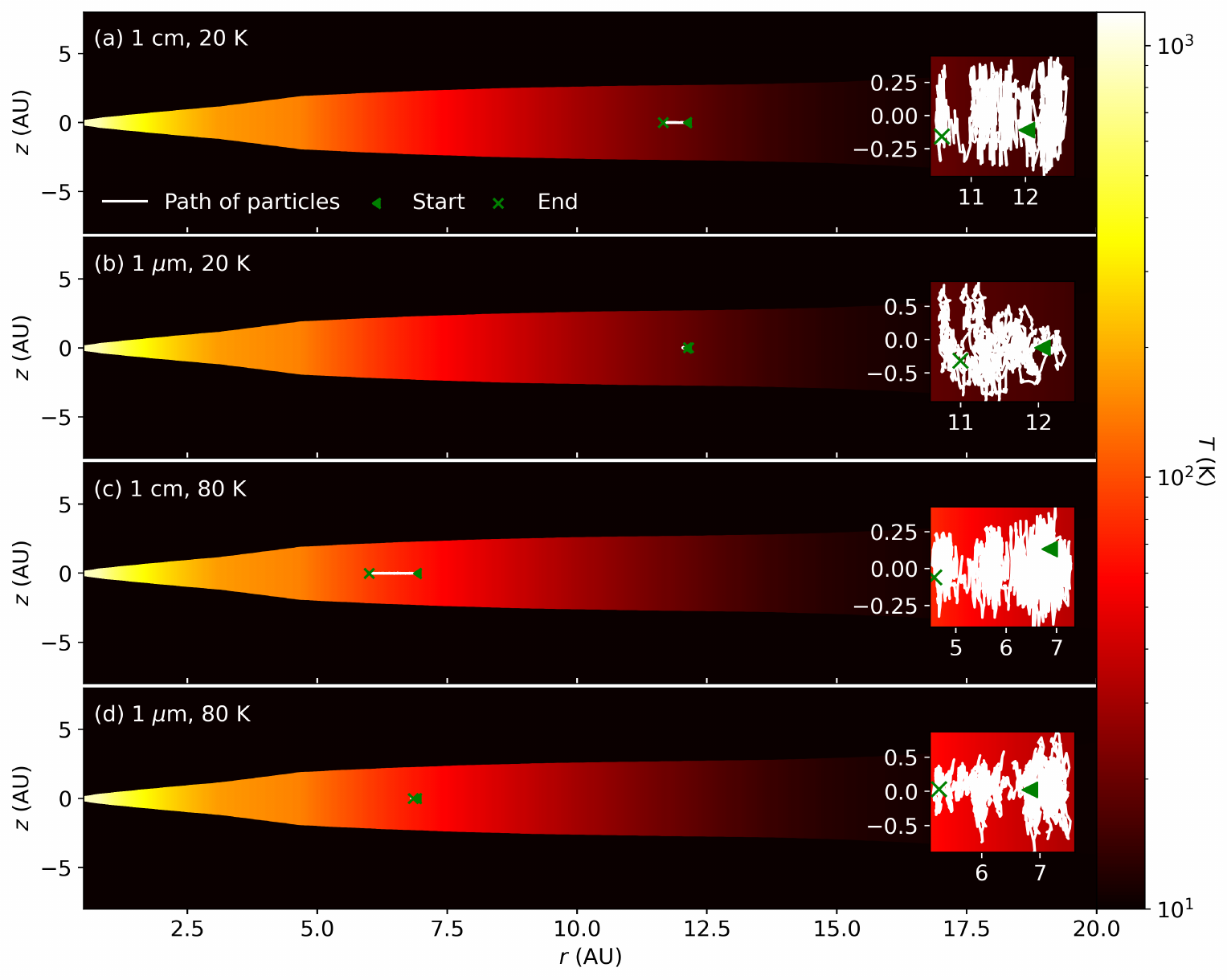}
\caption{Two-dimensional temperature map of the disk after 280 kyr of evolution. The averaged trajectories of particles over this period, plotted along the $r$ and $z$ axes, are shown in white. The panels present the trajectories of particles with sizes of 1 cm and 1 $\mu$m, released at initial disk temperatures of 20 K (panels a and b) and 80 K (panels c and d). Green triangles indicate the initial positions of the particles, while green crosses mark their final positions. In each panel, the detailed trajectory of an individual particle is highlighted within the square on the right.}
\label{fig:2d_profile}
\end{figure*}

Figure \ref{fig:2d_profile} shows a two-dimensional temperature profile of the disk at t = 280 kyr of its evolution. The mean particle trajectory, averaged over 500 particles, is superimposed to illustrate the general transport pattern. For comparison, the trajectory of a single particle is shown in the inset on the right, highlighting deviations from the averaged trend. Simulations were performed for particles of 1 cm and 1 $\mu$m size, initially released at 12 AU and 7 AU, with local temperatures of about 20 K and 80 K respectively. The range of possible particles' trajectories is detailed in appendix \ref{appendix:radial_traj}.

\begin{figure*}
\centering
\includegraphics[width=\textwidth]{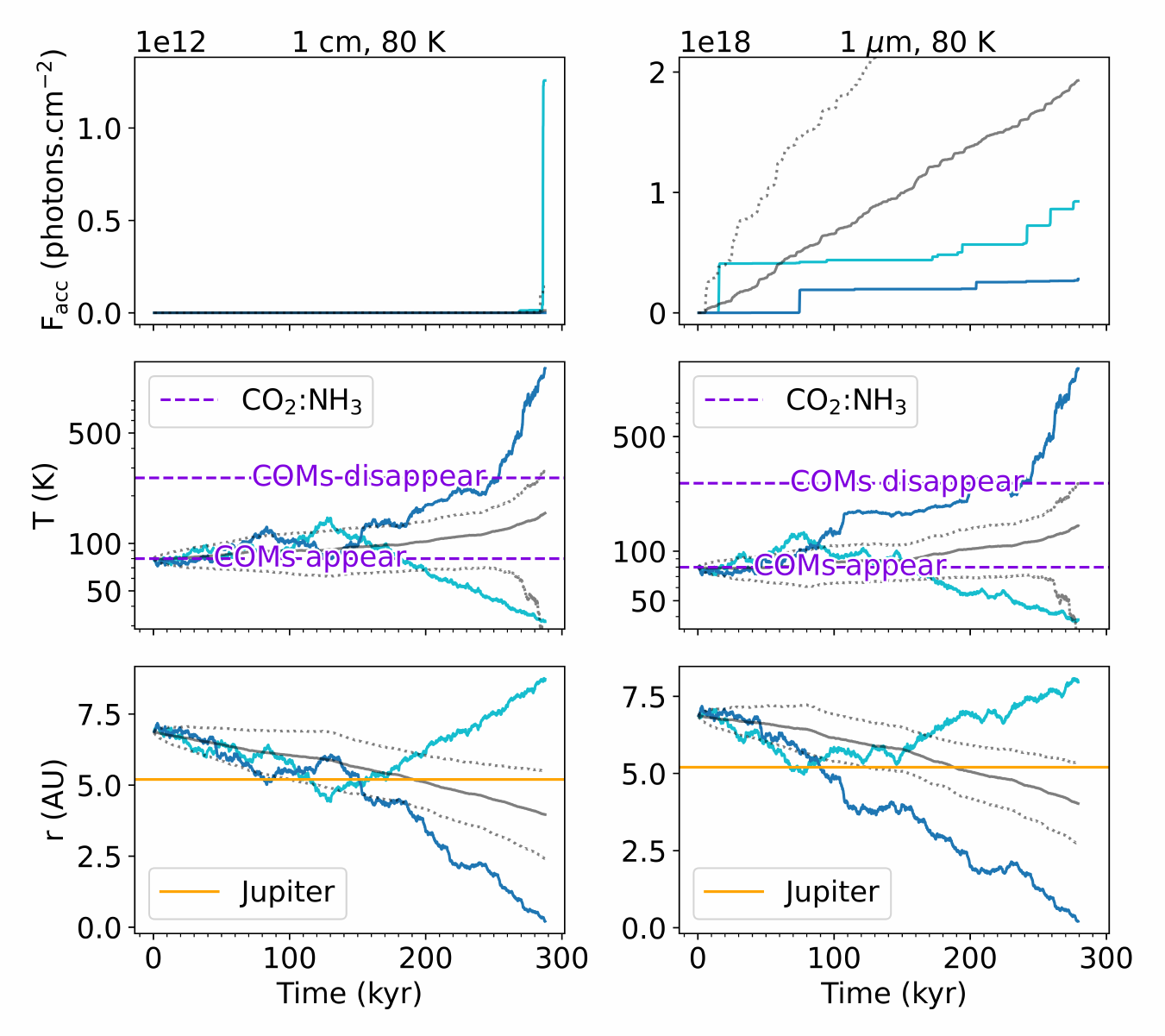}
\caption{Trajectories and irradiation conditions of selected individual particles, 1 cm (left column) and 1 $\mu$m (right column), released at 7 AU, with a PSN temperature of 80 K, during 290 and 280 kyr of PSN evolution, respectively. The particles were selected from those that succeeded in forming COMs and delivering them to Jupiter's orbit. The gray solid line shows the average accumulated irradiation, temperature, and trajectories of these particles, and the gray dotted line shows the corresponding standard deviation. Light and dark blue lines represent the accumulated irradiation, temperature, and radial trajectories of the particles that end up farthest and closest to the Sun, respectively. Top panels: Irradiation accumulated by these particles along their trajectory through the disk. Middle panels: Temperature encountered by the same particles, with temperature ranges for COM formation and destruction indicated by dashed horizontal lines, based on experimental data \citep{bossa_carbamic_2008}. Bottom panels: Radial trajectories of the same particles, with the location of the Jupiter system marked by the horizontal line.}
\label{fig:80K_COMs_trajectory}
\end{figure*}

\begin{figure*}
\centering
\includegraphics[width=\textwidth]{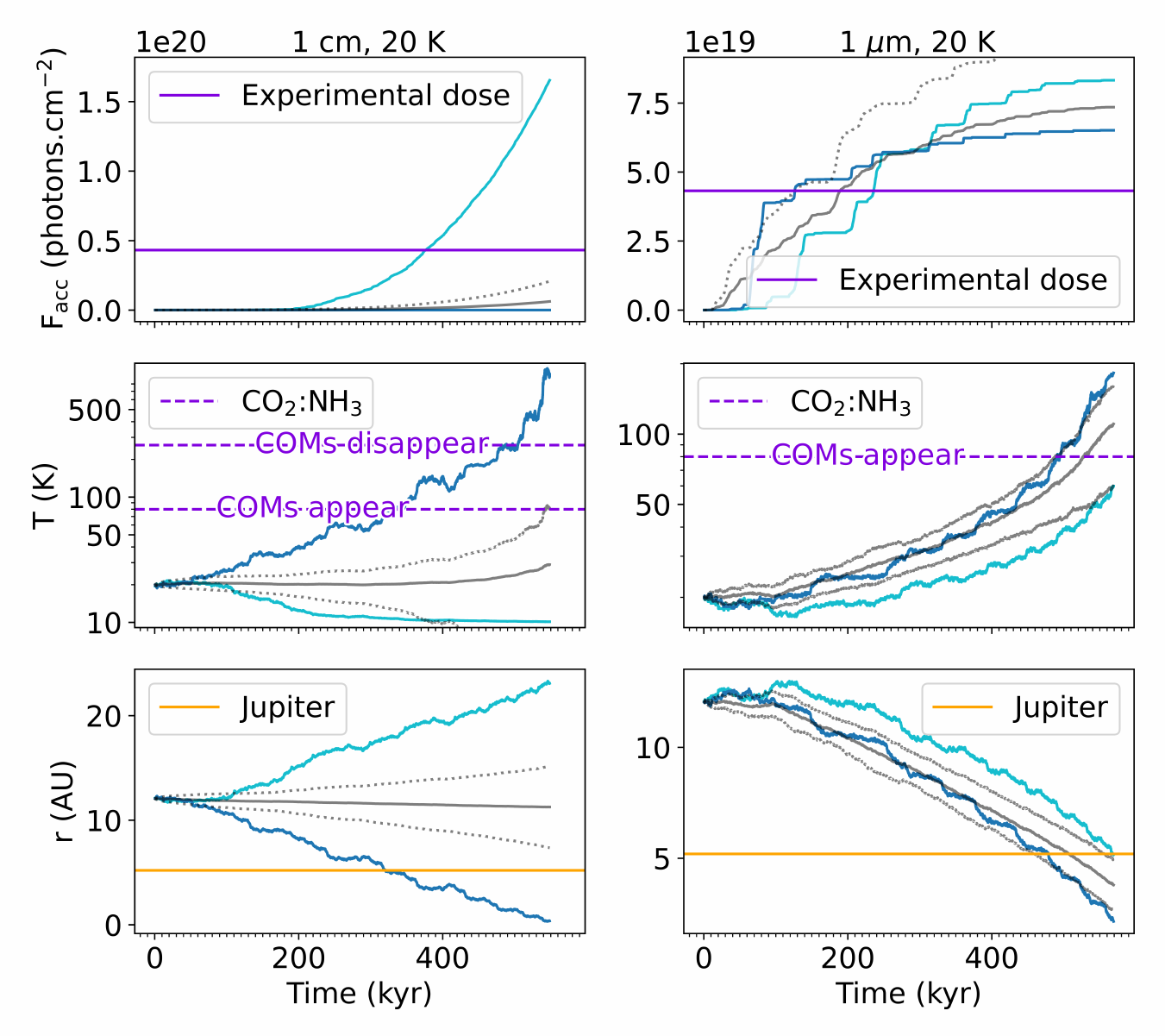}
\caption{Same as Fig. \ref{fig:80K_COMs_trajectory}, but for particles released at 12 AU (PSN temperature = 20 K), with simulation durations of 550 kyr (left) and 570 kyr (right). In this case, $\sim$2\% of the 500 simulated particles of 1 $\mu$m form COMs and reach the formation region of the Galilean moons, while none of the 1 cm particles do. Thus, the left panel shows all the 1 cm particles released at 20 K.}
\label{fig:20K_COMs_trajectory}
\end{figure*}

The results highlight the strong influence of particle size on vertical and radial transport. As averaging smooths out vertical variations, these are analyzed through individual particle trajectories. 1 $\mu$m particles released from 12 AU reach heights up to 1.6 AU, while 1 cm particles stay closer to the midplane, with altitudes not exceeding 0.63 AU. On average, 1 $\mu$m particles reach the end of their radial trajectories at 12.1 AU from the Sun, while 1 cm particles terminate their trajectories at 11.6 AU after 280 kyr of PSN evolution. A similar trend is observed for particles released from 7 AU. The 1 $\mu$m particles reach maximum altitudes of up to 1.3 AU, whereas 1 cm particles are constrained to altitudes no greater than 0.57 AU. Radially, the 1 $\mu$m particles complete their trajectories at about 6.8 AU, while the 1 cm particles settle closer to the star at 5.98 AU after 280 kyr of PSN evolution, on average. However, both particle sizes --larger ones driven by radial drift and smaller ones influenced by gas velocity -- tend to migrate faster when released closer to the star. In particular, micrometric particles tightly coupled to the gas exhibit faster migration within the centrifugal radius, closely mirroring the gas behavior. Finally, as shown in Figs. \ref{fig:20K_COMs_trajectory} and \ref{fig:80K_COMs_trajectory}, both 1 $\mu$m and 1 cm particles can drift outward. Since the radial trajectory of the particle is also driven by a random term (the last term in equation \ref{eq:euler,x}), the consequence is that outliers particles will mainly drift outward.

Figure \ref{fig:80K_COMs_trajectory} shows the accumulated irradiation $F_{\text{acc}}$, temperature $T$, and radial trajectories $r$ of the individual 1 cm and 1 $\mu$m particles that ultimately reach the farthest and closest points to the Sun. These particles were selected from the group that successfully transported COMs to the system of Jupiter. The particles are initially released at 7 AU, corresponding to a PSN temperature of 80 K. The top panels show the irradiation accumulated by the particles along their trajectories compared to the experimental irradiation threshold for COM formation from NH$_3$:CO$_2$ ices (4.32 × 10$^{19}$ photons cm$^{-2}$). The middle panels show the temperatures encountered by the same particles, along with the experimental temperatures for COM formation and destruction by thermal processing. The region of the disk where the temperature is between 80 and 260 K is referred to as the COMs thermal stability zone. COMs are formed at temperatures above 80 K, but they disappear when the temperature exceeds 260 K. The bottom panels show the trajectories of these particles relative to the position of the system of Jupiter (about 5.2 AU).
Particles close to the star are in a warm, low-irradiation region, while those farther away are in a cold, high-irradiation environment. To successfully deliver COMs to the system of Jupiter, these particles must either accumulate an irradiation dose equal to or exceeding the experimental threshold, or reach the COM-forming temperature and then travel to the vicinity of Jupiter. Simulations of 1 cm and 1 $\mu$m particles were conducted over 290 kyr and 280 kyr of PSN evolution, respectively.

The 1 cm particles released at 80 K form COMs, as shown in Figure \ref{fig:80K_COMs_trajectory}. However, COM formation occurs exclusively by thermal processing, since none of these particles accumulate the experimental irradiation dose of 4.32 × 10$^{19}$ photons cm$^{-2}$ necessary for COMs formation. Notably, these particles accumulate less than $1.5 \times 10^{12}$ photons cm$^{-2}$ over this period. These particles experience COM formation from heating within the first kyr after release in the PSN. As shown in Fig. \ref{fig:80K_COMs_trajectory}, the particle that ultimately ends its trajectory closest to the Sun crosses the orbit of the system of Jupiter at approximately 80 kyr and 150 kyr of PSN evolution. Similarly, the particle that drifts farthest from the Sun at the end of the simulation reaches the orbit of the Jupiter system between about 110 kyr and 171 kyr of PSN evolution.
The closest particle reaches COM sublimation temperatures at 253 kyr of PSN evolution, well after crossing Jupiter's orbit. In contrast, the most distant particle drifts outward and never experiences such heating. Overall, 95.6\% of the 500 simulated particles undergo COM formation, although only 44.4\% reach the location of the system of Jupiter.

Figure \ref{fig:80K_COMs_trajectory} also shows that smaller 1 $\mu$m particles released at 80 K are capable of forming COMs and transporting them to the Jupiter region. While 94 \% of the 500 simulated particles undergo COM formation, only 29.6 \% are delivered to the system of Jupiter after 280 kyr of PSN evolution. Similar to the larger particles, almost none of these smaller particles form COMs through irradiation, as they do not accumulate more than $2.4 \times 10^{19}$ photons cm$^{-2}$ over 280 kyr of PSN evolution, which is below the experimental irradiation dose of $4.32 \times 10^{19}$. As a result, COM formation occurs solely through thermal processing, within the first dozen kyr of PSN evolution after particle release. In addition, the closest particle reaches temperatures as high as 260 K, causing the destruction of the COMs after 200 kyr of PSN evolution. The COMs carried by the particles reach the system of Jupiter between 57.8 and 280 kyr of PSN evolution.

Figure \ref{fig:20K_COMs_trajectory} presents the accumulated irradiation dose $F_{\text{acc}}$, temperature $T$, and radial trajectories $r$ of individual 1 cm and 1 $\mu$m particles that ultimately reach the farthest and closest distances from the Sun. Here, particles are released at 12 AU (PSN temperature of 20 K), and simulations of 1 cm and 1 $\mu$m particles extend over 550 kyr and 570 kyr, respectively. The 1 $\mu$m particles shown are a subset of those that successfully deliver COMs to the system of Jupiter, whereas all 1 cm particles are included, as none contribute to COM delivery. The Figure illustrates the challenge for 1 cm and 1 $\mu$m particles released at $\sim$12 AU to form COMs and be successfully transported to the Jupiter system region within the first 0.5 Myr of PSN evolution. In fact, none of the 1 cm particles succeed, although 4.6\% form COMs by thermal processing and 3\% form COMs by UV irradiation. The particle closest to the Sun at the end of the simulations reaches the Jupiter system location after 320 kyr of PSN evolution, but COM formation occurs later. In particular, this particle enters the COM thermal stability zone after 323 kyr of PSN evolution and never accumulates more than $7.6 \times 10^{15}$ photons cm$^{-2}$, a value four orders of magnitude below the threshold. The most distant particle never reaches either the Jupiter system location or the COM thermal stability zone, although it reaches the experimental irradiation dose of $4.32 \times 10^{19}$ photons cm$^{-2}$, and thus forms COMs only after 376 kyr of PSN evolution.

In the case of the 1 $\mu$m particles, almost 80\% of the 500 simulated particles form COMs due to UV irradiation, while about 1.4\% form COMs due to heating. However, only 2\% of the particles successfully transport COMs to the system of Jupiter. Among these, the particle closest to the Sun at the end of the simulations reaches the irradiation threshold at 150 kyr and arrives at Jupiter at 433 kyr. In contrast, the most distant particle reaches the irradiation threshold at 255 kyr and arrives at Jupiter at 511 kyr. The higher irradiation of the closest particle, especially in the early stages of the PSN, is mainly due to the vertical component of its trajectory, which is not shown in the figures. Vertical turbulence, represented in our model by the random term in Eq. \ref{eq:euler,z}, can cause some particles to reach high altitudes, where they experience more irradiation, while others remain confined to the midplane. Notably, the closest particle reaches the temperature required for thermal processing of the COMs at 454 kyr, 200 kyr after the formation of COMs from UV irradiation. Since this particle reaches the irradiation threshold earlier, the ice it carries has already been converted into COMs by the time it reaches the COM formation temperature. This is an illustration of the fact that earlier irradiation may allow the formation of COMs before the conditions necessary for thermal processing are reached.

\section{Discussion and Conclusion} \label{sec:discussion}

In this study, we show that N--bearing COMs can form in the PSN and be transported into Jupiter’s orbit, assuming the planet formed at its present position. For these COMs to be incorporated into the Galilean moons, they must first be transported to the CPD when they reach 5.2 AU. This assumption overlooks potential challenges in particle delivery, such as gap formation around the growing proto-Jupiter. Furthermore, the integrity of COMs during their transport from the PSN to the CPD remains uncertain, as some scenarios suggest that these structures could be destroyed in the process. Desorption during entry, which could lead to COM destruction and recondensation into simpler ices within the CPD \citep{Oberg2022, Oberg2023}, is not considered here. Furthermore, studies have suggested that material accreted onto the CPD may pass through the surface layers of the PPD at high altitudes, where it is exposed to enhanced irradiation \citep{morbidelli_meridional_2014,szulagyi_accretion_2014}. Such irradiation can alter the molecular composition, promoting the formation of a larger number of COMs and, ultimately, their destruction. However, other studies \citep{Oberg2022,Oberg2023} do not rule out the possibility that at least part of the material could remain preserved during its transport to the CPD.

Within the CPD, COMs may be further degraded by UV irradiation or heating. Specifically, a more massive CPD could expose COMs to elevated temperatures, causing sublimation and rendering them unsuitable for moon formation \citep{Schneeberger2025}. To mitigate this, we assume a cold CPD environment and rapid accretion by protomoons to minimize irradiation exposure. This assumption is supported by the supersolar D/H ratio measured in Callisto \citep{Clark2019}, which suggests that the Galilean moons' building blocks likely originated unaltered from the PSN. This, in turn, implies that at least Callisto formed in a cold environment \citep{Horner2008,Nimmo25}. This conclusion could be extended to the satellites of Saturn, whose D/H has been measured to be supersolar \citep{WaiteJr2009,brown_deuterated_2025}. This reinforces the hypothesis of the conservation of material from the PPD to the CPD, and is consistent with the hypothesis of cold CDPs.

A key factor is the heating of the disk’s surface by the central star. This heating raises the temperature in the upper layers, where micrometer-sized particles could, in principle, experience elevated temperatures during their passage at high altitudes. However, using the same model as in \citet{okamoto_effects_2024}, we find that, within the regions and epochs considered, 1 $\mu$m particles do not reach these thermally enhanced layers. The same occurs with 1 cm particles, which cannot reach altitudes as high as 1 $\mu$m particles.

Our study also shows that while COM formation is not inherently complex, their transport to Jupiter’s orbit and eventual accretion by the Galilean moons presents significant challenges. Particles released beyond $\sim$12 AU readily reach the irradiation threshold for COM formation within a few hundred kyr. Our simulations indicate that 80\% of micrometer-sized particles released at 12 AU successfully form COMs through UV irradiation of NH$_3$:CO$_2$ ices. However, most of these particles fail to reach Jupiter’s region due to their large initial heliocentric distance and slow inward drift.

Conversely, particles released within $\sim$7 AU are less likely to receive sufficient irradiation for COM formation but are more prone to thermal processing due to the higher temperatures in the inner PSN. For example, about 95\% of these particles manage to form COMs. However, less than 45\% of the centimeter-sized particles and less than 30\% of the micrometer-sized particles released at 7 AU deliver COMs to the Jupiter region. The remaining COM--bearing particles fail to reach Jupiter, either because they form COMs closer to the star than Jupiter and never drift outwards, or because they drift outwards after COM formation, preventing them from reaching Jupiter. Overall, larger particles tend to favor thermal processing, since they drift inward faster and cannot reach the irradiated top layers of the disk.

Scenarios suggesting the formation of Jupiter in the 3 AU region of the PSN \citep{Walsh2011, Bitsch2015} or within the first 100 kyr of its evolution \citep{Zhu2012} are challenged by the abundance profile of N--bearing molecules. The icelines of N$_2$ and NH$_3$, the primary nitrogen carriers, are initially located beyond the Jovian system. Our simulations show that the NH$_3$ iceline, with a condensation temperature of 75 K \citep{Fray_Schmitt_2009}, starts beyond Jupiter's orbit and takes 160 kyr to migrate inwards to its current position. The N$_2$ iceline, which condenses at $\sim$19 K \citep{Fray_Schmitt_2009}, takes even longer to migrate. Therefore, the presence of nitrogen in Jupiter's CPD is hardly consistent with scenarios where Jupiter formed closer to the Sun or in the early stages of the PSN. However, our study shows that nitrogen can be delivered to the moons via alternative pathways, such as COMs formed in the PSN via the thermal and UV processing of NH$_3$:CO$_2$ ices. Furthermore, alternative processes could also account for the delivery of nitrogen to the Jovian system. For instance, the formation of nitrogen-bearing refractory organic compounds in the colder regions of the PSN or in the parent molecular cloud could have enabled the transport of nitrogen toward the hotter inner regions.

Providing a mechanism for the incorporation of nitrogen--bearing COMs into the building blocks of the Galilean moons is crucial, as these compounds may have played a key role in the evolution of their hydrospheres, originally thought to be global oceans in equilibrium with primordial atmospheres \citep{lunine_clathrate_1987,lunine_massive_1992}. First, NH$_3$ profoundly affects the distribution of volatiles in the primordial hydrospheres of icy moons. When NH$_3$ reacts with CO$_2$, a volatile thought to be present in Europa's ocean \citep{trumbo_distribution_2023, villanueva_endogenous_2023}, the resulting chemical equilibrium affects both the pH of the ocean and the abundance of volatiles in the atmosphere. In particular, interactions between CO$_2$ and NH$_3$ speciation products lead to the formation of carbonates and NH$_2$COO$^-$ ions. The concentrations of these ions, together with the pH of the ocean, strongly depend on the CO$_2$/NH$_3$ ratio supplied during moon formation \citep{amsler_moulanier_role_2025}. Second, as demonstrated in this study, the Galilean moons may have directly accreted complex nitrogen-bearing species such as NH$_2$COO$^-$. If preserved during the moons' accretion, these compounds could influence CO$_2$--NH$_3$ chemical equilibrium, impacting ocean pH and habitability conditions. For example, a high NH$_3$/CO$_2$ ratio would make the ocean more alkaline, favoring CO$_2$ sequestration as carbonates. This in turn would reduce the availability of free CO$_2$ in the ocean, limiting its transport to the ice surface. Such processes are particularly relevant given the observed presence of CO$_2$ on the surfaces of Europa, Ganymede, and Callisto \citep{McCord_1998, trumbo_distribution_2023, villanueva_endogenous_2023}. 

Finally, this study relies on several key assumptions that may be refined in the future:

\begin{itemize}
\item \textit{Restricted set of formation and destruction processes:} Our model does not account for hydrogenation \citep{Linnartz2011} or secondary UV irradiation induced by cosmic rays interacting with H$_2$ gas \citep{Prasad_Tarafdar_1983}. In addition, the study of reactions involving other molecules and the integration of a chemical photoreaction network \citep{Takehara2022,Ochiai2024} could provide a more comprehensive understanding of the chemical evolution in PPDs. Although we compared the total irradiation dose in the PPD with experimental conditions, the much lower irradiation rate in the PPD is likely to limit the molecular diversity. This limitation arises because radicals formed under such conditions have more time to recombine into their precursors, reducing the efficiency of complex molecule formation.

\item \textit{Limits on opacity and UV irradiation flux:} The mean Rosseland opacity used in this study is not fully suited for UV irradiation \citep{benest2024}, since it is frequency-averaged. The \cite{Birnstiel2018} opacity model, more appropriate for our nominal UV frequency range, predicts lower particle irradiation but assumes a constant value across the PSN, without taking opacity regimes into account, unlike temperature-- and pressure--dependent opacities. An interstellar irradiation flux with G$_0$ = 1 was assumed, though studies suggest variability up to G$_0$ = 10$^4$ \citep{Adams_Proszkow_Fatuzzo_Myers_2006, Yeghikyan_2009}, significantly increasing irradiation doses.

\item \textit{Influence of PPD parameters:} Variation of parameters such as the $\alpha$ viscosity, mass accretion rate, and particle properties could affect results \citep{benest2024}.

\item \textit{Migration timescale and growth of particles:} The simulated particle trajectories in this work span timescales between 280 and 550 kyr. However, the actual lifetimes of the particles are likely shorter, as they may fragment through collisions or grow into bigger objects before completing their trajectories. For instance, the growth timescale of particles is estimated to be less than 10 kyr during the early stages of PSN evolution and approximately 100 kyr after 1 Myr of evolution in the regions considered \citep{brauer_coagulation_2008,birnstiel_simple_2012,Aguichine_Mousis_Devouard_Ronnet_2020,benest2024}. Particle growth, which was neglected in our model, has been shown to influence trajectories and fragmentation \citep{Misener_Krijt_Ciesla_2019}. For these reasons, accounting for particle growth in future studies could yield valuable insights.

\item \textit{Jupiter’s Formation and Migration:} The formation of Jupiter is assumed to have occurred in situ at 5.2 AU. However, dynamical models propose a scenario involving inward formation followed by outward migration \citep{Walsh2011}. This alternative pathway would likely diminish irradiation--driven COM accretion during the inward phase but would increase thermal-processing--driven COM accretion during the subsequent outward migration.

\item \textit{Timing of CPD Formation:} Jupiter may have formed within the first 100 kyr of PSN evolution if gravitational instability governed its formation \citep{Zhu2012, Aguichine22, Schneeberger_Mousis_Aguichine_Lunine_2023}. In this scenario, the formation of its surrounding CPD during the final stage of planetary growth would be consistent with the accretion of material, including COMs, before 500 kyr of PSN evolution. Alternatively, models advocating for Jupiter’s later formation via pebble accretion \citep{Alibert2018} require a later accretion of material onto its CPD. At this stage, the NH$_3$ iceline remains closer to the Sun than Jupiter, facilitating the incorporation of N-bearing ices into the CPD.

\item \textit{Formation conditions of the moons:} During their accretion phase, the Galilean moons may have experienced high temperatures, with conditions depending on moon size, impactor size distribution, and accretion timescale \citep{Stevenson1986, Canup_Ward_2002,Bierson_Nimmo_2020,bennacer_conditions_2025}. Several studies suggest that Io and Europa underwent high-temperature accretion \citep{Bierson_Nimmo_2020}, leading to the destruction of COMs. However, recent findings indicate that Europa may have instead formed under low-temperature conditions \citep{Petricca2025}. Ganymede’s accretion could have been either warm or cold, depending on the model considered \citep{Canup_Ward_2002,Bierson_Nimmo_2020,Bjonnes2022,bennacer_conditions_2025}, while Callisto is generally believed to have formed under cold accretion conditions \citep{Canup_Ward_2002, Bierson_Nimmo_2020, Bjonnes2022,bennacer_conditions_2025}. Consequently, depending on their accretion history, the icy Galilean moons may have retained some of their primordial COMs.  

\end{itemize}

Our conclusions about the amount of COMs delivered to the Galilean moons would be affected if these processes were included in our model.

\section*{Acknowledgements}

The project leading to this publication has received funding from the Excellence Initiative of Aix-Marseille Universit\'e--A*Midex, a French ``Investissements d’Avenir program'' AMX-21-IET-018. This research holds as part of the project FACOM (ANR-22-CE49-0005-01\_ACT) and has benefited from a funding provided by l'Agence Nationale de la Recherche (ANR) under the Generic Call for Proposals 2022. We thank Yannis Bennacer and Antoine Schneeberger for fruitful discussions during the preparation of this paper.

\section*{Data Availability}

The data will be made available upon reasonable request to the
authors.



\bibliographystyle{mnras}
\bibliography{Tom_Aliz} 




\appendix

\section{Radial particle trajectories}  \label{appendix:radial_traj}

\begin{figure}
    \centering
    \includegraphics[width=\linewidth]{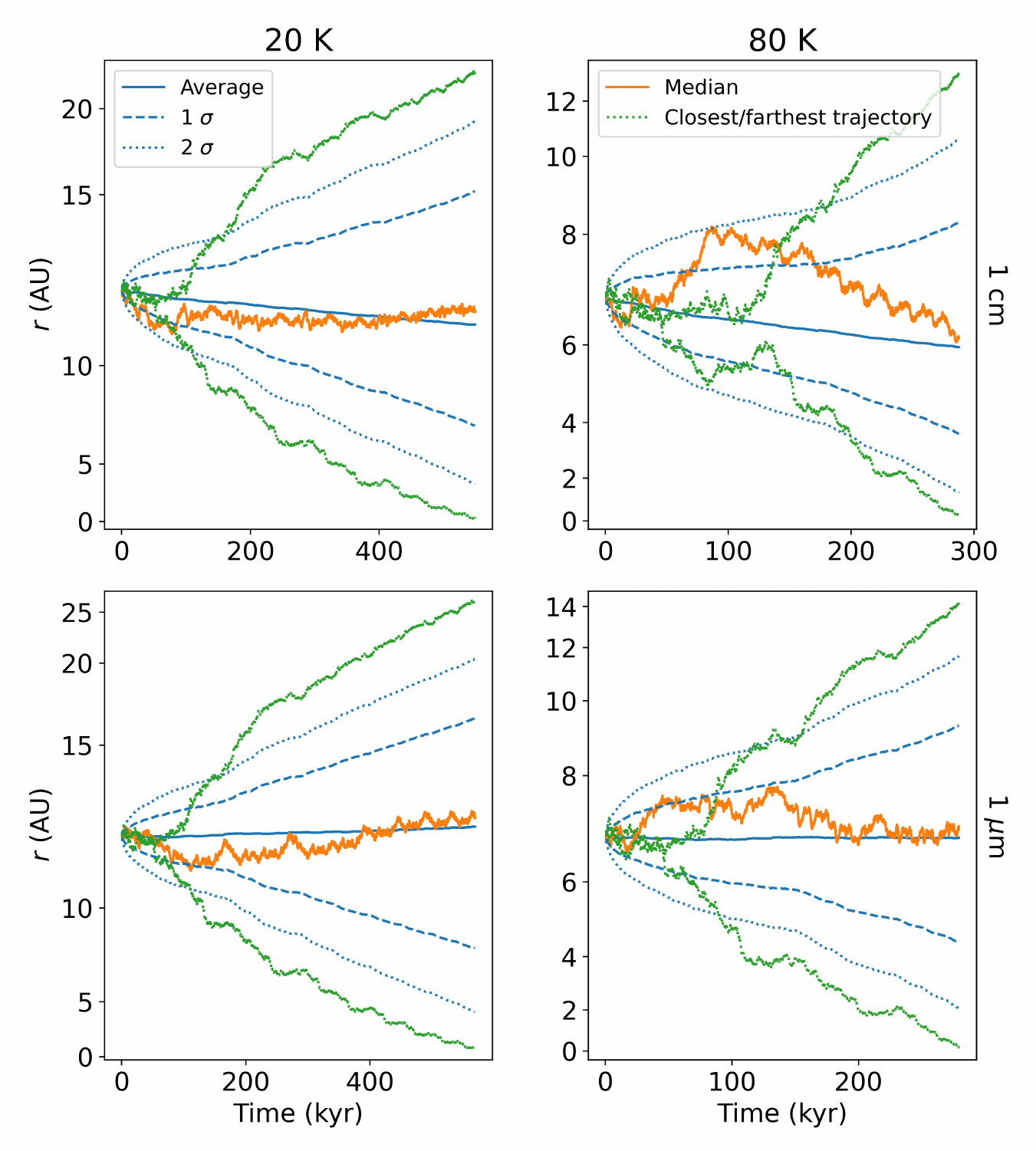}
    \caption{Radial trajectories of 1 cm (top panels) and 1 $\mu$m (bottom panels) particles released at 20 K (left) and 80 K (right). The mean trajectory is shown as a solid blue line, with the corresponding 1-$\sigma$ and 2-$\sigma$ dispersions represented by dashed and dotted blue lines, respectively. The median trajectory is shown in orange, while the green dotted lines indicate the innermost and outermost trajectories relative to the star.}
    \label{fig:r_stats}
\end{figure}

Figure \ref{fig:r_stats} shows the average trajectory for each case, along with the corresponding 1-$\sigma$ and 2-$\sigma$ dispersions. The median trajectory and the two extreme trajectories are also displayed. On average, the 1 cm particles exhibit inward drift.

\bsp	
\label{lastpage}
\end{document}